\documentclass[prd,reprint,showpacs,showkeys]{revtex4}
\usepackage{amsfonts}
\usepackage{amssymb}
\usepackage{amsmath}
\usepackage{graphicx}
\usepackage[font={footnotesize,it}]{caption}

\begin{document}

\title{Cycloid Experiment for freshmen physics labs}
\author{R. Akoglu}
\email{resat.akoglu@emu.edu.tr}
\author{S. Habib Mazharimousavi}
\email{habib.mazhari@emu.edu.tr}
\author{M. Halilsoy}
\email{mustafa.halilsoy@emu.edu.tr}
\affiliation{Department of Physics, Eastern Mediterranean University, Gazima\~{g}usa,
Turkey. }
\date{\today }

\begin{abstract}
We establish an instructive experiment to investigate the minimum time curve
traveled by a small billiard ball rolling in a grooved track under gravity.
Our intention is to popularize the concept of \textit{minimum time curve}
anew, and to propose it as a feasible physics experiment both for freshmen
and sophomore classes. We observed that even the non-physics major students
did enjoy such a cycloid experiment.
\end{abstract}

\maketitle

\address{Department of Physics, Eastern Mediterranean University,\\
G. Magosa, north Cyprus, Mersin-10, Turkey\\
}

\section{\textbf{The minimum time of descent under gravity}}

The minimum time of descent under gravity has historical importance in
connection with Fermat's principle, a problem that remains ever popular to
the readers of general physics matters \cite{1,2,3}. Our aim here is to
propose an experiment for the introductory mechanics laboratory such that
the students explore the minimum time curve known as a cycloid (Fig. 1)
themselves. A small billiard ball rolls from rest under gravity from an
initial (fixed) point O to a final (fixed) point A along different paths
(Fig. 2 and 3). The relation between its speed $v$ and vertical position $y$
can easily be found from the energy conservation i.e. $T+V=const.$ or $%
\Delta T=-\Delta V$ in which $T$ and $V$ are the kinetic and the potential
energy respectively.

Out of infinite number of possible paths joining O to A we are interested in
the one that takes the minimum time. This is one of the typical extremal
problems encountered in mechanics \cite{4} under the title of \textit{%
Brachistochrone problem} whose solution is given in almost all books of
mechanics. The time of slide between O and A is given by 
\begin{equation}
\Delta t=\int_{o}^{A}\frac{ds}{v}
\end{equation}%
in which $v=\sqrt{2gy}$ and $ds$ is the element of the arclength along the
path (Eq. (2) below). Note that for a billiard ball, as an extended object
with inertia the relation between $v$ and $y$ modifies into $v=\sqrt{\frac{10%
}{7}gy}$, which doesn't change the nature of the minimum time curve. We
shall state simply the result: The curve is a cycloid expressed
mathematically in parametric form 
\begin{gather}
x=a\left( \varphi -\sin \varphi \right)   \notag \\
y=a\left( 1-\cos \varphi \right)  \\
\left( 0\leq \varphi \leq 2\pi \right)   \notag
\end{gather}%
where $2a$, is the maximum point $y_{\max }$ since $y$ is a downward
coordinate along the curve. For different paths the pathlength of the curve $%
S_{OA}$ can be obtained from the integral expression 
\begin{equation}
S_{OA}=\int_{O}^{A}ds=\int_{\varphi =0}^{\varphi =\pi }\sqrt{dx^{2}+dy^{2}}.
\end{equation}%
Mathematically it is not possible to evaluate this arclenght unless we know
the exact equation for the curve. Two exceptional cases, are the straight
line and the cycloid. As a curve the cycloid has the property that at O/A it
becomes tangent to the vertical/ horizontal axis. Although the lower point
(i.e. A) could be chosen anywhere before the tangent point is reached, for
experimental purpose we deliberately employ the half cycloid, so that
identification of the Brachistochrone becomes simpler. By using a string and
ruler we can measure each pathlength to great accuracy. The experimental
data will enable us to identify the minimum time curve, namely the cycloid.

\section{\textbf{Apparatus and Experiment}}

1) A thin, grooved track made of a long flexible metal bar (or hard plastic)
fixed by clamps.

2) A small billiard ball.

3) A digital timer connected to a fork-type light barrier.

4) String and ruler to measure arclengths.

The experimental set up is seen in Fig. 2. We note that the track must be at
least 2 meters long both for a good demonstration and to detect significant
time differences. The track is fixed at A by a screw while the other end of
the track passing through the fixed point O is variable This gives us the
freedom to test different paths, with the crucial requirement that in each
case the starting point O at which the timer is triggered electronically
remains fixed. This particular point is the most sensitive part of the
experiment which is overcome by using a fork-type light barrier (optic eyes)
both at O and A. As the path varies the light flash can be tolerated to
intersect any point of the ball with a negligible error. Let us add also
that an extra piece of track at A is necessary to provide proper flattening
at the minimum of the inverted half cycloid. From Fig. 3, path $P_{1}$ is
identified as a straight line which is added here for comparison with the
otherwise curved paths. As we change the path down from $P_{1}$ to $P_{8}$
we record the time of each descent by a digital timer. We observe that as we
go from $P_{1}$ to $P_{4}$ with exact tangential touches to the axes, the
time decreases, reaching a minimum at $P_{4}$. From $P_{5}$ on, the time
starts to increase again toward $P_{8}$ with almost tangential touches at A.
In this way we verify experimentally that $P_{4}$ can be identified, as the
minimum time curve. By using a string and ruler we measure the length of
each path as soon as we record its time of descent. The length $S_{1}$ of
the straight line path $P_{1}$, for example will be ( recall from Eq. (2))
from the simple hypotenuse theorem

%
\begin{figure}[h]
\includegraphics[width=120mm,scale=0.7]{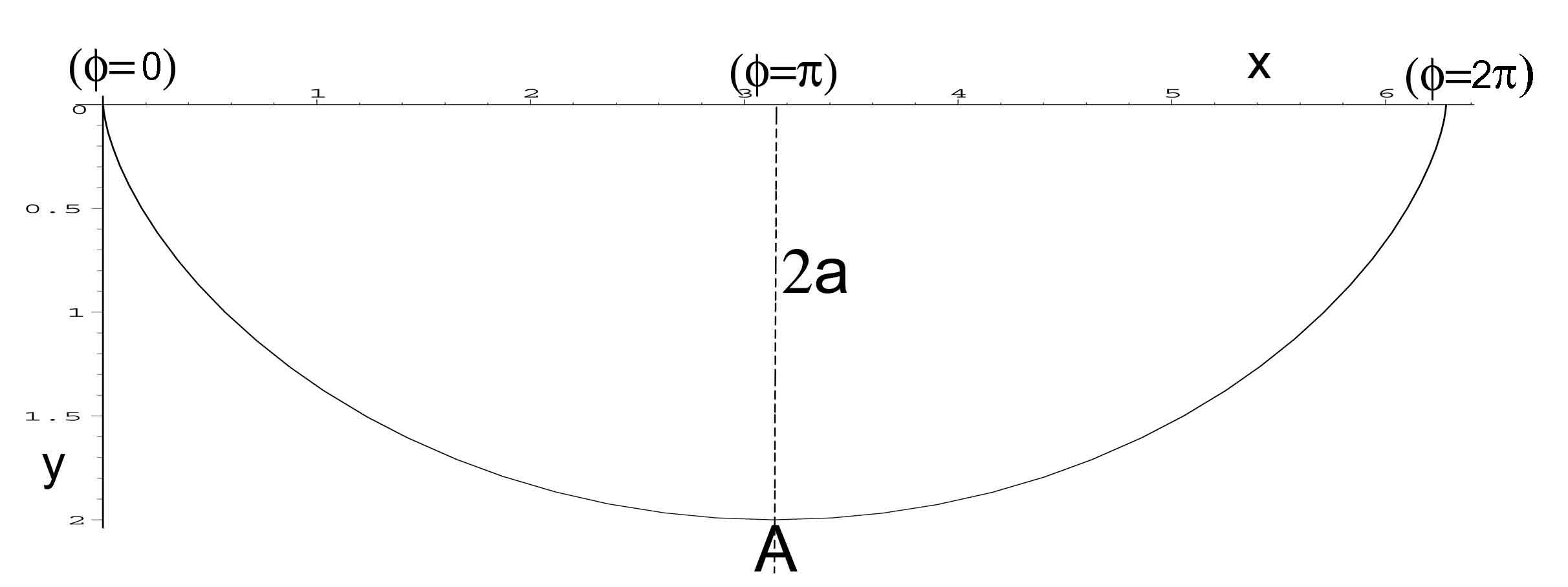}
\caption{A cycle in an inverted cycloid for $0\leq \protect\phi \leq 2%
\protect\pi ,$ with maximum height $2a$. }
\label{fig: 1}
\end{figure}
\begin{figure}[h]
\includegraphics[width=120mm,scale=0.7]{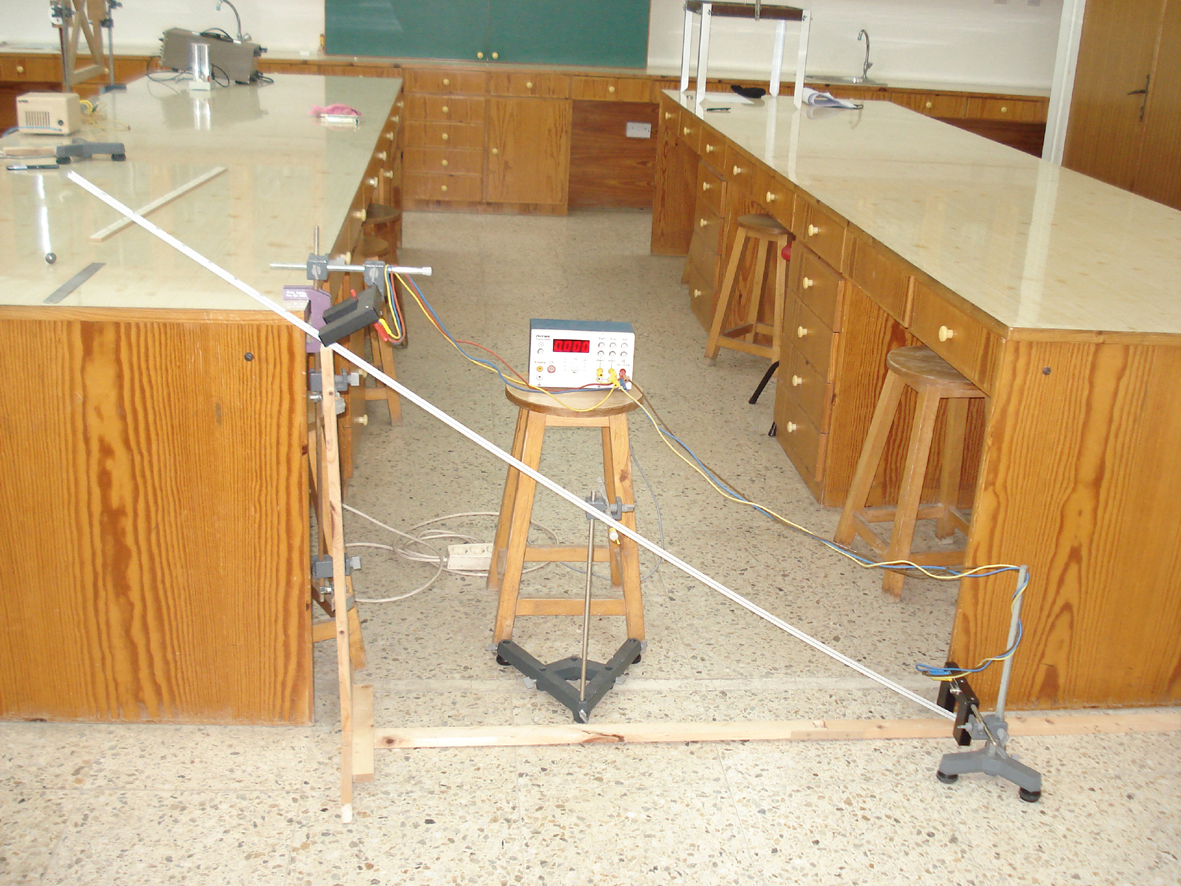}
\caption{Our complete experimental apparatus / setup in display.}
\label{Pic: 2}
\end{figure}
\begin{figure}[h]
\includegraphics[width=120mm,scale=0.7]{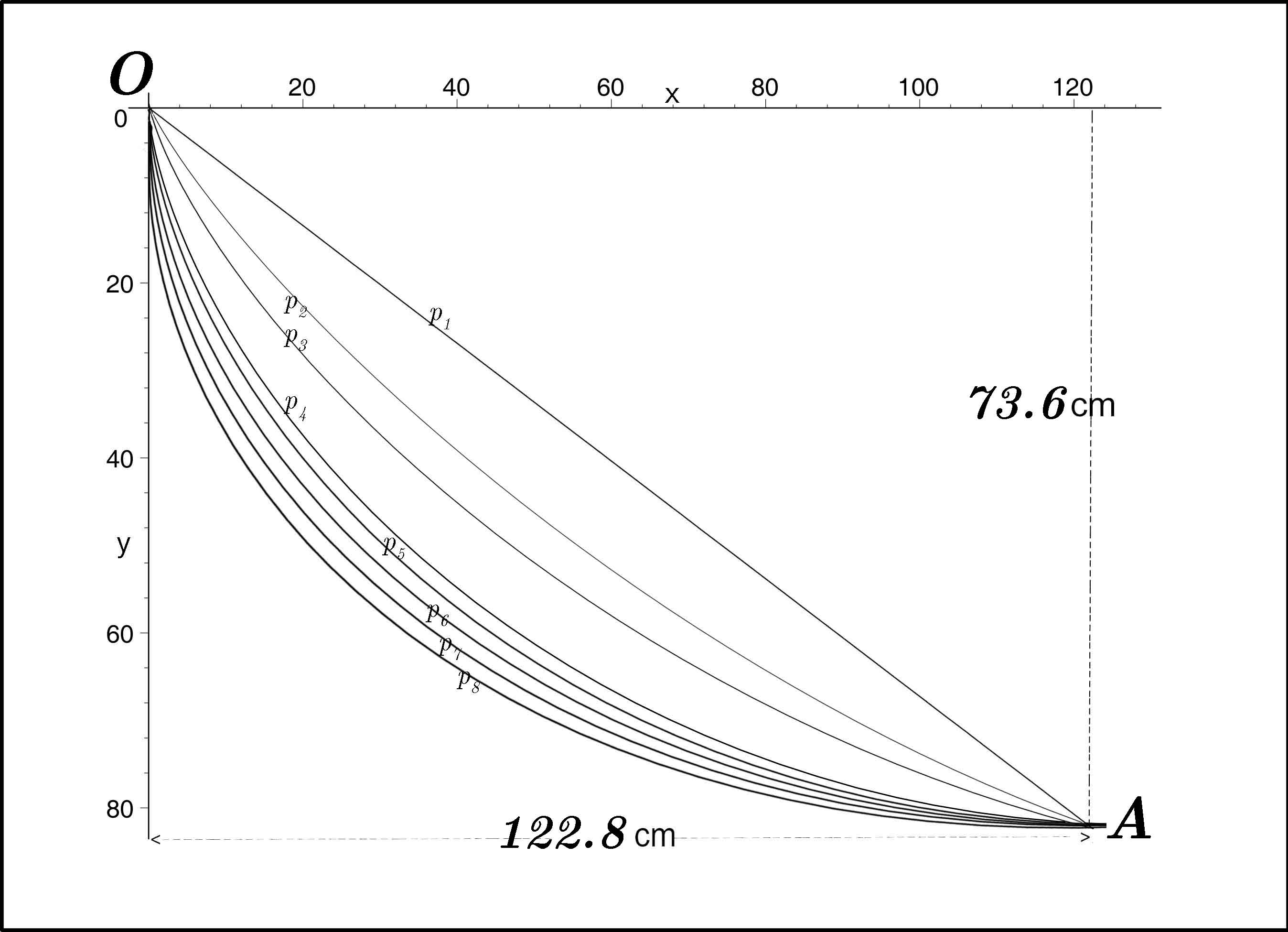}
\caption{Rolling of a billiard ball from $O$ to $A$. We consider $8$
different paths, labelled as $P_{1},P_{2}...P_{8}.$ }
\label{fig: 2}
\end{figure}
\begin{equation}
S_{1}=\sqrt{x^{2}+y^{2}}=2a\sqrt{1+\left( \frac{\pi }{2}\right) ^{2}}
\end{equation}%
where $2a=73.4\ cm$ stands for the maximum height. From Fig. 1 and Tab. 1 we
see that $S_{1}$ is calculated theoretically ($\approx 136.5$ $cm$) and
experimentally ($\approx 137.1$ $cm$), is acceptable within the limits of
error analysis. Addition of errors involved in the readings of arclengths,
time and averaging results will minimize the differences. It should also be
taken care that while in rolling, the ball doesn't distort the track. \ \ \
\ \ \ \ \ \ \ \ \ \ \ \ \ \ \ \ \ \ \ \ \ \ 

\begin{table}[th]
\caption{The length and the time taken, corresponding to each path}
\label{table:nonlin}
\centering
\begin{tabular}{ccccccccc}
\hline\hline
Path & $P_{1}$ & $P_{2}$ & $P_{3}$ & $P_{4}$ & $P_{5}$ & $P_{6}$ & $P_{7}$ & 
$P_{8}$ \\[1ex] \hline
Length $S_{i}$ $(cm)$ & $137.1$ & $142.1$ & $144.1$ & $146.2$ & $150.2$ & $%
154.3$ & $158.5$ & $162.6$ \\ 
Time $(sec)$ & $0.740$ & $0.663$ & $0.654$ & $0.652$ & $0.654$ & $0.656$ & $%
0.668$ & $0.677$ \\ \hline
\end{tabular}
\end{table}

The pathlength of the cycloid i. e., Eq. (3) by substitution from Eq. (1)
can be obtained to satisfy%
\begin{equation}
S_{4}=4a.
\end{equation}%
Experimentally all one has to do after taking each time record is to check
that the minimum time curve satisfies Eq. (5), and it is tangent at O/A
which characterize nothing but the Brachistochrone problem. Theoretically we
have $S_{4}=146.8cm$ while experimental value is $146.2$ $cm$ which implies
an error less than one percent.

As an alternative method which we tried also to convince ourselves, we
suggest to use a digital camera to take the picture of each path and locate
them on a common paper for comparison. As noticed, in performing the
experiment we have used only half of the cycloid $0\leq \varphi \leq \pi .$
If space is available a longer track can be used to cover the second half , $%
\pi \leq \varphi \leq 2\pi $, as well. Owing to the symmetry of a cycloid,
however, this is not necessary at all.

\section{Conclusion}

A cycloid arises in many aspects of life. It is the curve generated by a
fixed point on the rim of a circle rolling on a straight line. Diving of
birds / jet fighters toward their targets, watery sliding platforms in aqua
parks are some of the examples in which minimum time curves and therefore
cycloids are involved. In comparison with a circle and ellipse, cycloid is a
less familiar curve at the introductory level of mathematics / geometry. The
unusual nature comes from the fact that both the angle $\varphi $ and its
trigonometric function arise together so that the angle can't be inverted in
terms of coordinates in easy terms. Yet the details of mathematics which are
more apt for the sophomore classes can easily be suppressed. Changing the
track each time before rolling the ball, measuring both time of fall and
length of the curve are easy and much instructive to conduct as a physics
experiment. Main task the students are supposed to do is to fill the data in
Table 1. It will not be difficult for students to explore that cycloid is
truly the minimum time curve of fall under constant gravitational field. Let
us complete our analysis by connecting that simple extension of our
experiment can be done by using variable initial points. Namely, instead of
the fixed point O the ball can be released from any other point between O
and A which doesn't change the time of fall. This introduces the students
with the problem of tautochrone, which is also interesting.

\end{document}